\documentclass[prl,twocolumn,showpacs,nofootinbibs,superscriptaddress
]{revtex4}
\usepackage{epsf}
\usepackage{epsfig}
\usepackage{amsmath}

\def\beq{\begin{equation}}
\def\eeq{\end{equation}}
\def\bea{\begin{eqnarray}}
\def\eea{\end{eqnarray}}
\def\beqa{\begin{equation}\begin{array}{l}}
\def\eeqa{\end{array}\end{equation}}

\begin{document}
\preprint{WM-07-109}
\preprint{JLAB-THY-07-}

\title{
Empirical transverse charge densities in the nucleon and 
the nucleon-to-$\Delta$ transition}

\author{Carl E. Carlson}
\affiliation{Physics Department, College of William and Mary,
Williamsburg, VA 23187, USA}

\author{Marc Vanderhaeghen}
\affiliation{Physics Department, College of William and Mary,
Williamsburg, VA 23187, USA}
\affiliation{Theory Center, Thomas Jefferson National Accelerator Facility, 
Newport News, VA 23606, USA}

\date{October 2, 2007}

\begin{abstract}
Using only the current empirical information on the nucleon electromagnetic 
form factors we map out the transverse charge density in proton and neutron   
as viewed from a light front moving 
towards a transversely polarized nucleon. These charge densities are 
characterized by a dipole pattern, in addition to the monopole field 
corresponding with the unpolarized density. 
Furthermore, we use the latest empirical 
information on the $N \to \Delta$ transition form factors to map out the 
transition charge density which induces the $N \to \Delta$ excitation. 
This transition charge density in a transversely polarized $N$ 
and $\Delta$ contains both monopole, dipole and quadrupole patterns, 
the latter corresponding with a deformation of the hadron's charge 
distribution.  
\end{abstract}

\pacs{13.40.Gp, 14.20.Dh, 14.20.Gk}

\maketitle
\thispagestyle{empty}

Electromagnetic form factors (FFs) of the nucleon
are the standard source of information
on the nucleon structure and as such have been studied extensively; for 
recent reviews see {\it e.g.}  
Refs.~\cite{HydeWright:2004gh,Arrington:2006zm,Perdrisat:2006hj}. 
The FFs describing the transition of the nucleon to its first
excited state, $\Delta$(1232), 
contain complementary information, such as the sensitivity
on the nucleon shape; see Ref.~\cite{Pascalutsa:2006up} for a recent review. 
\newline
\indent
In more recent years, generalized parton distributions (GPDs) have 
been discussed (see {\it e.g.}  
Refs.~\cite{Ji:1998pc,Goeke:2001tz,Diehl:2003ny,Belitsky:2005qn} for some 
reviews) as a tool to access the distribution of partons in the transverse 
plane~\cite{Burkardt:2000za}, and first calculations of these spatial  
distributions have been performed within lattice QCD~\cite{LHPC:2003is} 
and hadronic models (see {\it e.g.}~\cite{Pasquini:2007xz} for a recent
evaluation). By integrating the GPDs over all parton 
momentum fractions, they reduce to FFs. 
Given the large amount of precise data on FFs it is of 
interest to exhibit directly the spatial information which 
results from these data. This has been done recently 
in Ref.~\cite{Miller:2007uy} 
for an unpolarized nucleon. In this Letter we extend that 
work to the case of a transversely polarized nucleon as well as to 
map out the transition charge density which induces 
the $N \to \Delta$ excitation. 

In the following we consider the electromagnetic (e.m.) $N \to N$ and 
$N \to \Delta$ transitions when viewed from a light front moving towards 
the baryon. Equivalently, this corresponds with 
a frame where the baryons have a 
large momentum-component along the $z$-axis chosen along the direction of 
$P = (p + p^\prime)/2$, where $p$ ($p^\prime$) are the intial (final) 
baryon four-momenta. We indicate the baryon light-front + component by $P^+$ 
(defining $a^\pm \equiv a^0 \pm a^3$). 
We can furthermore choose a symmetric frame where the virtual photon 
four-momentum $q$ has $q^+ = 0$, 
and has a transverse component (lying in the $xy$-plane)
indicated by the transverse vector $\vec q_\perp$, satsifying 
$q^2 = - {\vec q_\perp}^{\, 2} \equiv - Q^2$. 
In such a symmetric frame, the virtual photon only couples to forward moving 
partons and the + component of the 
electromagnetic current $J^+$ 
has the interpretation of the quark charge density operator. It is  
given by~: $J^+(0) = +2/3 \, \bar u(0) \gamma^+ u(0) - 1/3 \, 
\bar d(0) \gamma^+(0) d(0)$, considering only $u$ and $d$ quarks. 
Each term in the expression is a positive operator since 
$\bar q \gamma^+ q \propto | \gamma^+ q |^2$. 

\begin{figure}[h]
\begin{center}
\includegraphics[width =7.7cm]{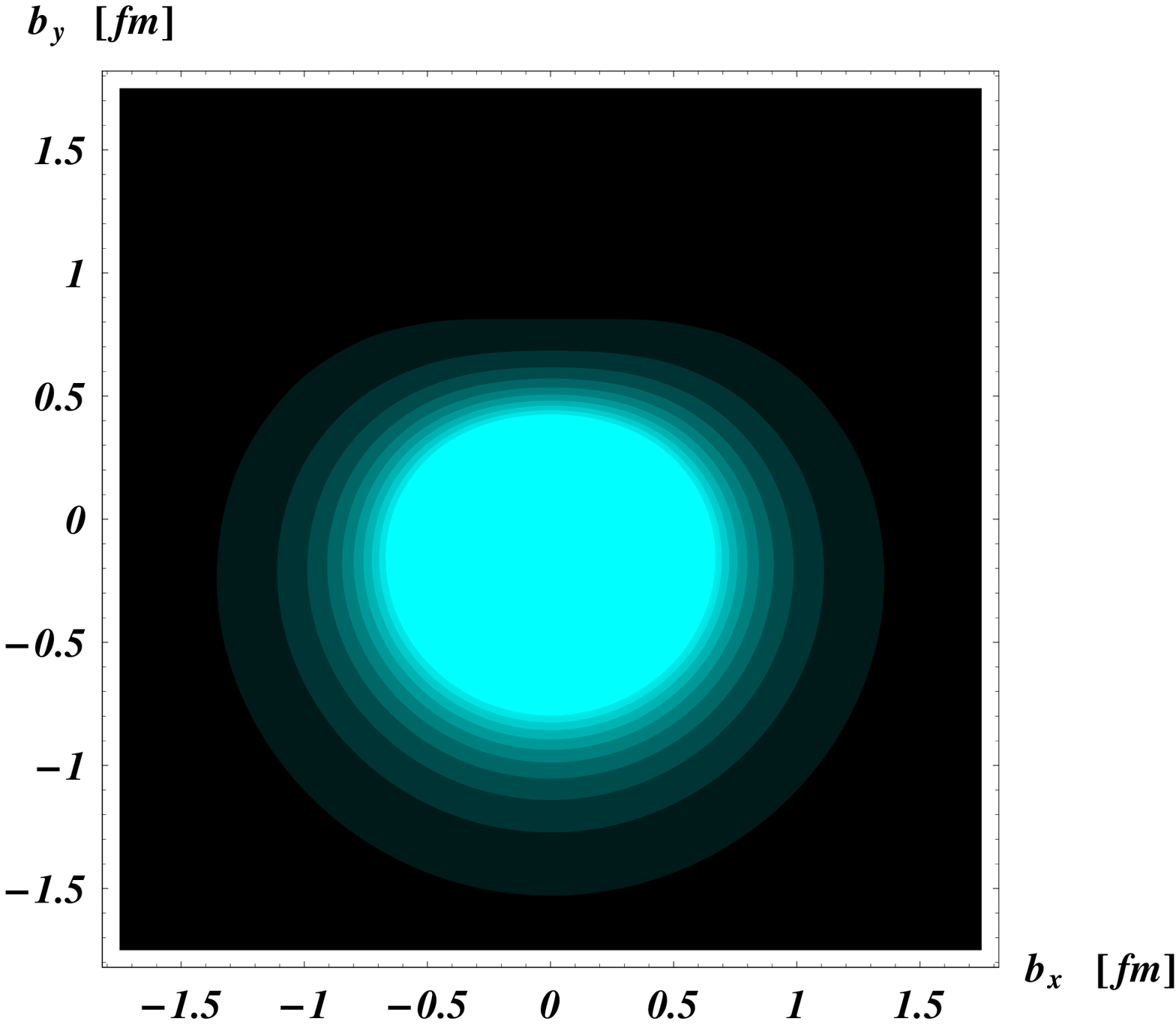}
\end{center}
\hspace{.5cm}
\includegraphics[width =7.5cm]{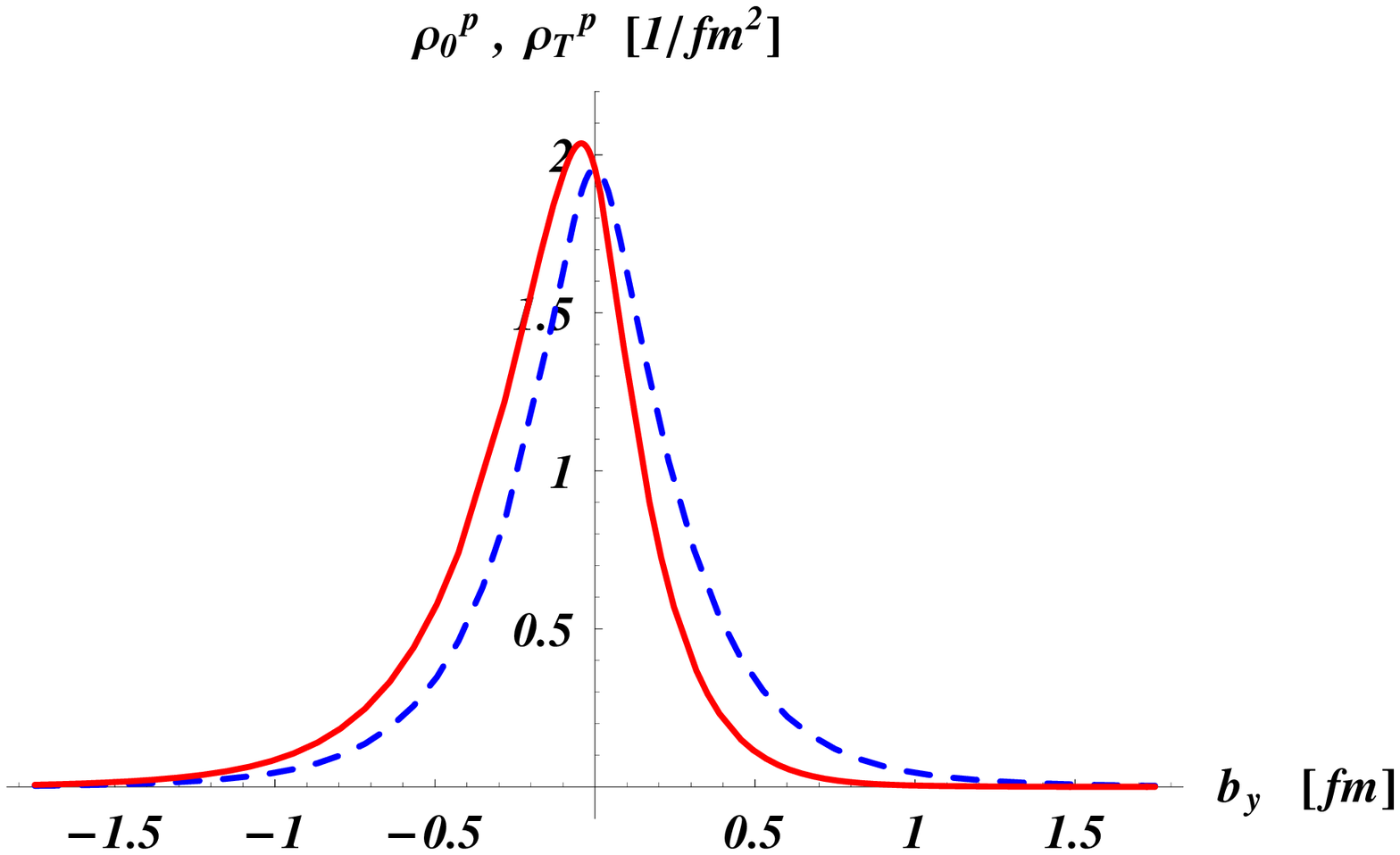}
\caption{Quark transverse charge densities in the {\it proton}. 
The upper panel shows the density in the transverse plane for a 
proton polarized along the $x$-axis. The light (dark) regions correspond with
largest (smallest) values of the density. 
The lower panel compares the density along the $y$-axis 
for an unpolarized proton (dashed curve), 
and for a proton polarized along the $x$-axis (solid curve). 
For the proton e.m. FFs, we use the empirical parameterization of 
Arrington {\it et al.}~\cite{Arrington:2007ux}. }
\label{fig:proton}
\end{figure}

Following \cite{Burkardt:2000za, Miller:2007uy}, one can then 
define quark transverse charge densities in a nucleon as~:
\begin{eqnarray}
\rho_0^N(\vec b) &\equiv& \int \frac{d^2 \vec q_\perp}{(2 \pi)^2} \,  
e^{i \, \vec q \cdot \vec b} \, \frac{1}{2 P^+} 
\label{eq:ndens1} \\
&& \hspace{1.25cm}\times  
\langle P^+, \frac{\vec q_\perp}{2}, \lambda \,|\, J^+(0) \,|\, 
P^+, -\frac{\vec q_\perp}{2}, \lambda  \rangle, \nonumber
\end{eqnarray}
where the 2-dimensional vector $\vec b$ denotes the position (in the 
$xy$-plane) from the transverse {\it c.m.} of the nucleon, and  
$\lambda = \pm 1/2$ denotes the nucleon (light-front) helicity. 
\newline
\indent
Using the Dirac $F_1$ nucleon e.m. FF, Eq.~(\ref{eq:ndens1}) can be 
expressed as~\cite{Miller:2007uy}~:  
\begin{eqnarray}
\rho_0^N(b) = \int_0^\infty \frac{d Q}{2 \pi} Q \, J_0(b \, Q) F_1(Q^2), 
\label{eq:ndens2}
\end{eqnarray}
with $J_n$ denotes the cylindrical Bessel function of order $n$. 
Note that $\rho_0^N$ only depends on $b = |\vec b|$. 
It has the interpretation of a quark charge density in the transverse plane 
for an unpolarized nucleon, and is well defined for all values of $b$, 
even when $b$ is smaller than the nucleon Compton wavelength. In contrast, the 
usual 3-dimensional Fourier transform of the matrix elements of $J^\mu$ in the 
Breit-frame (parameterized in terms of the Sachs FFs) 
becomes intrinsically ambiguous~\cite{Kelly:2002if}, 
due to the Lorentz contraction of the nucleon along its direction of motion.  
Although this does not affect the densities at larger distances (typically
larger than about 0.5 fm) the value for the densities at smaller densities
is merely a reflection of the prescription how to relate the experimentally 
measured Sachs FFs at large $Q^2$ 
with the intrinsic charge and magnetization FFs. A feature of viewing 
the nucleon when ``riding a photon'' is that one  
gets rid of the longitudinal direction. 
This allows one to project the charge density 
(in the case of the $J^+$ operator) 
on the transverse plane, which does not get Lorentz contracted. 
In this way, it was found {\it e.g.} in Ref.~\cite{Miller:2007uy} that
the neutron charge density reveals the well known negative contribution 
at large distances, around 1.5~fm, 
due to the pion cloud, a positive contribution at 
intermediate $b$ values, and a negative core at $b$ values smaller than 
about 0.3~fm, reflecting the large $Q^2$ behavior of the neutron Dirac FF. 
\newline
\indent
It was shown in Ref.~\cite{Burkardt:2000za} that 
one can also define a probability distribution to find a quark 
with a given momentum fraction $x$ of $P^+$, and at a 
given transverse position $b$ in the nucleon, 
when considering a nucleon polarized in the $xy$-direction. 
In the following, we denote this transverse polarization direction by 
$\vec S_\perp = \cos \phi_S \hat e_x + \sin \phi_S \hat e_y$. 
When integrating the resulting GPD, depending on $x$ and $\vec b$, 
over all values of $x$, one can define a quark charge density in the 
transverse plane for a transversely polarized nucleon as~:
\begin{eqnarray}
\rho_T^N(\vec b) &\equiv& \int \frac{d^2 \vec q_\perp}{(2 \pi)^2} \,
e^{i \, \vec q \cdot \vec b} \, \frac{1}{2 P^+} 
\label{eq:ndens3} \\
&& \hspace{-.75cm}\times  
\langle P^+, \frac{\vec q_\perp}{2}, s_\perp = +\frac{1}{2} 
\,|\, J^+(0) \,|\, 
P^+, -\frac{\vec q_\perp}{2}, s_\perp = +\frac{1}{2}  \rangle, \nonumber 
\end{eqnarray}
where $s_\perp = +1/2$ is the nucleon 
spin projection along the direction of $\vec S_\perp$. 
By working out the Fourier transform in Eq.~(\ref{eq:ndens3}), one obtains~:
\begin{eqnarray}
\rho_T^N(\vec b) &=& \rho_0^N(b) 
\label{eq:ndens4} \\ 
&-& \sin (\phi_b - \phi_S) \, 
\int_0^\infty \frac{d Q}{2 \pi} \frac{Q^2}{2 M_N} \, J_1(b \, Q) 
F_2(Q^2), \nonumber  
\end{eqnarray}
where the second term, which describes the deviation from the circular
symmetric unpolarized charge density, depends on the orientation of 
$\vec b = b ( \cos \phi_b \hat e_x + \sin \phi_b \hat e_y )$. 
Furthermore, this term 
depends on the Pauli FF $F_2$ and the nucleon mass $M_N$. 
\newline
\indent
In the following we are using the current empirical information on the nucleon 
e.m. FFs to extract the transverse charge density in a transversely polarized 
nucleon, complementing the pictures given in Ref.~\cite{Miller:2007uy}, 
for the transverse charge densities in an unpolarized nucleon. 
For the proton e.m. FFs, we use the recent empirical parameterization of 
Ref.~\cite{Arrington:2007ux} and show the resulting transverse charge density
for a proton polarized along the $x$-axis ({\it i.e.} for $\phi_S = 0$) 
in Fig.~\ref{fig:proton}. 
One notices from Fig.~\ref{fig:proton} that 
polarizing the proton along the $x$-axis 
leads to an induced electric dipole moment along the 
negative $y$-axis which is equal to the value of the 
anomalous magnetic moment, {\it i.e.} $F_2(0)$ (in units $e/2 M_N$) as first 
noticed in Ref.~\cite{Burkardt:2000za}. One can understand this induced 
electric dipole field pattern based on the classic work of 
Ref.~\cite{Einstein:1908} (see also the pedagogical explanation in
Ref.~\cite{Krotkov99}). The nucleon spin along the $x$-axis is the source 
of a magnetic dipole field, which we denote by $\vec B$. 
An observer moving towards the nucleon with velocity $\vec v$ will see an
electric dipole field pattern with $\vec E^\prime = - \gamma (\vec v \times
\vec B)$ giving rise to the observed effect.  

\begin{figure}[h]
\begin{center}
\includegraphics[width =7.7cm]{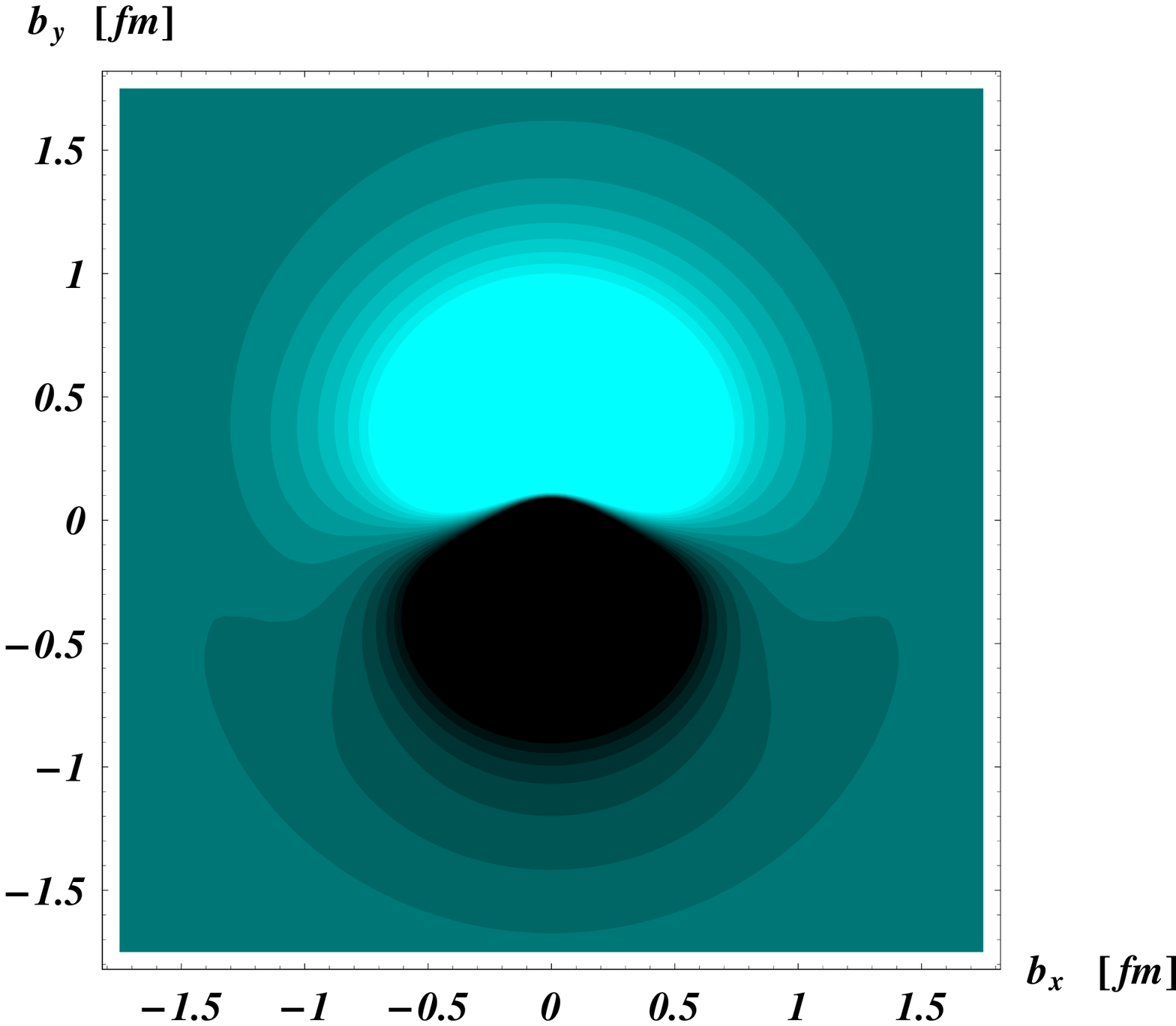}
\end{center}
\hspace{.5cm}
\includegraphics[width =7.5cm]{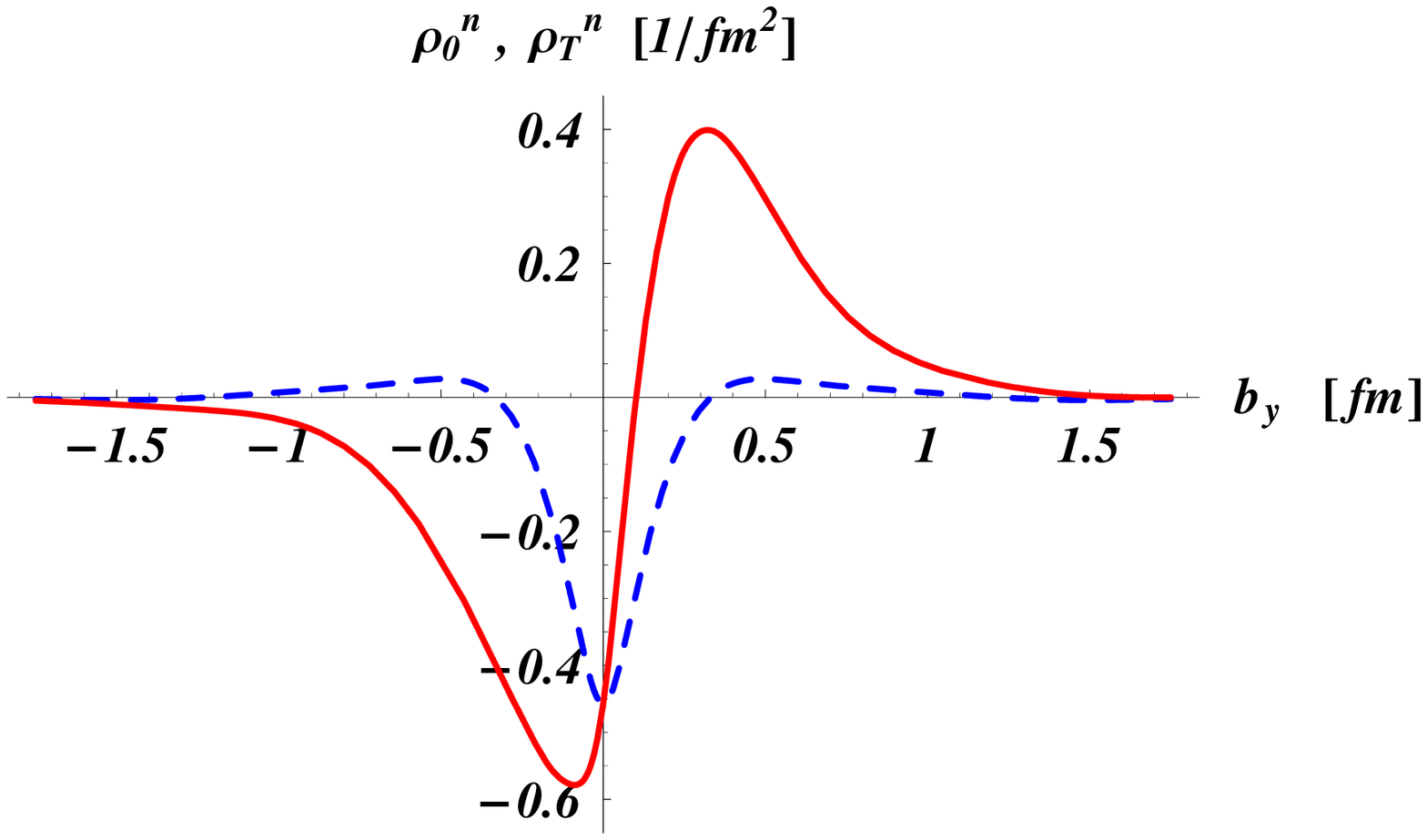}
\caption{Same as Fig.~\ref{fig:proton} for the 
quark transverse charge densities in the {\it neutron}. 
For the neutron e.m. FFs, we use the empirical parameterization of 
Bradford {\it et al.}~\cite{Bradford:2006yz}. }
\label{fig:neutron}
\end{figure}

For the neutron e.m. FF, we use the recent empirical parameterization of 
Ref.~\cite{Bradford:2006yz}. The corresponding transverse charge density 
for a neutron polarized along the $x$-axis is 
shown in Fig.~\ref{fig:neutron}. One notices that the neutron's 
unpolarized charge density gets displaced significantly due to the large 
(negative) value of the neutron anomalous magnetic moment,  
$F_{2n}(0) = -1.91$, which yields an induced electric dipole moment along the 
positive $y$-axis.  

We next generalize the above considerations to the 
$N \to \Delta$ {\it e.m.} transition as it allows access to 
$l = 2$ angular momentum components in the nucleon and/or $\Delta$ wave
functions. We will use the empirical
information on the $N \to \Delta$ transition FFs to study the 
quark transition charge densities in the transverse plane which 
induce the e.m. $N \to \Delta$ excitation.   
It is customary to characterize the three different 
types of the $\gamma N \Delta$ transitions 
in terms of the Jones--Scadron FFs $G^*_M$,  $G^*_E$,
$G^*_C$~\cite{Jones:1972ky}, corresponding with the magnetic dipole (M1), 
electric quadrupole (E2) and Coulomb quadrupole (C2) transitions respectively,
see Ref.~\cite{Pascalutsa:2006up} for details and definitions.   

We start by expressing the matrix elements of the $J^+(0)$ operator 
between $N$ and $\Delta$ states as~:  
\begin{eqnarray}
\langle P^+, \frac{\vec q_\perp}{2}, \lambda_\Delta | J^+(0) | 
P^+, -\frac{\vec q_\perp}{2}, \lambda_N  \rangle 
&=& (2 P^+)  e^{i  (\lambda_N - \lambda_\Delta) \phi_q} 
\nonumber \\
&\times& \, G^+_{\lambda_\Delta \, \lambda_N} (Q^2),
\label{eq:deldens1}
\end{eqnarray}
where $\lambda_N$ ($\lambda_\Delta$) denotes the nucleon ($\Delta$) light-front
helicities, 
and where $\vec q_\perp = Q ( \cos \phi_q \hat e_x + \sin \phi_q \hat e_y )$. 
Furthermore in Eq.~(\ref{eq:deldens1}), the helicity form factors 
$G^+_{\lambda_\Delta \, \lambda_N}$ depend on $Q^2$ only and can equivalently 
be expressed in terms of $G_M^\ast$, $G_E^\ast$, and $G_C^\ast$. 

We can then define a transition charge density for the unpolarized 
$N \to \Delta$ transition, which is given by the Fourier transform~:
\begin{eqnarray}
\rho_0^{N \Delta} (b) = \int_0^\infty \frac{d Q}{2 \pi} Q \, 
J_0(b \, Q) \, G^+_{+\frac{1}{2} \, +\frac{1}{2}} (Q^2),
\label{eq:deldens2}
\end{eqnarray}
where the helicity conserving $N \to \Delta$ FF 
$G^+_{+\frac{1}{2} \, +\frac{1}{2}}$ can be expressed in terms of 
$G_M^\ast$, $G_E^\ast$, and $G_C^\ast$ as~:
\begin{eqnarray}
&&G^+_{+\frac{1}{2} \, +\frac{1}{2}} = 
I \, \frac{(M_\Delta + M_N )}{M_N \, Q_+^2} \sqrt{\frac{3}{2}} 
\left( - \frac{Q^2}{4} \right)  \nonumber \\
&&\times \left\{ G_M^\ast + 
G_E^\ast \, \frac{3}{Q_-^2} \left[ (3 M_\Delta + M_N)(M_\Delta - M_N) - Q^2 
\right] \right. \nonumber \\
&&\hspace{.5cm} \left.  + \, 2 \, G_C^\ast   
\left[ -\frac{(M_\Delta + M_N)}{M_\Delta} + \frac{3 \, Q^2}{Q_-^2} 
\right] \right\} , 
\label{eq:deldens3}
\end{eqnarray}
with $M_\Delta = 1.232$~GeV the $\Delta$ mass,  
and where the isospin factor 
$I = \sqrt{2/3}$ for the $p \to \Delta^+$ transition, which we consider in all
of the following. We also introduced the shorthand notation 
$Q_\pm \equiv \sqrt{(M_\Delta \pm M_N)^2 +Q^2}$. 

The above unpolarized transition charge density gives us one combination of
the three independent $N \to \Delta$ FFs. To get information from 
the other combinations, we 
consider the transition charge densities for a transversely polarized 
$N$ and $\Delta$, both along the direction of $\vec S_\perp$ as~: 
\begin{eqnarray}
\rho_T^{N \Delta}(\vec b) &\equiv& \int \frac{d^2 \vec q_\perp}{(2 \pi)^2} \,
e^{i \, \vec q \cdot \vec b} \, \frac{1}{2 P^+} 
\label{eq:deldens4} \\
&& \hspace{-1cm}\times   
\langle P^+, \frac{\vec q_\perp}{2}, s^\Delta_\perp = +\frac{1}{2} 
\,|\, J^+(0) \,|\, 
P^+, -\frac{\vec q_\perp}{2}, s^N_\perp = +\frac{1}{2}  \rangle, \nonumber 
\end{eqnarray}
where $s^N_\perp = +1/2$ ($s^\Delta_\perp = +1/2$) are the nucleon ($\Delta$) 
spin projections along the direction of $\vec S_\perp$ respectively. 
By working out the Fourier transform in Eq.~(\ref{eq:deldens4}), one obtains~:
\begin{eqnarray}
\rho_T^{N \Delta} (\vec b) & = & \int_0^\infty \frac{d Q}{2 \pi} \, 
\frac{Q}{2}\, \left\{ 
J_0(b \, Q) \, G^+_{+\frac{1}{2} \, +\frac{1}{2}} 
\right. \nonumber \\  
&+& \sin(\phi_b - \phi_S) \, J_1(b \, Q) 
\left[ \sqrt{3} G^+_{+\frac{3}{2} \, +\frac{1}{2}}  
+ G^+_{+\frac{1}{2} \, -\frac{1}{2}} \right]
\nonumber \\
&-&\left.  \cos 2 (\phi_b - \phi_S) \, J_2(b \, Q) 
\sqrt{3} \, G^+_{+\frac{3}{2} \, -\frac{1}{2}}  
\right\} .
\label{eq:deldens5} 
\end{eqnarray}
One notices from Eq.~(\ref{eq:deldens5}) that besides half the unpolarized 
transition density, one obtains two more linearly independent structures. 
The $N \to \Delta$ FF combination with one unit of 
(light-front) helicity flip, which corresponds with a dipole field pattern in
the charge density, can be expressed in terms 
of $G_M^\ast$, $G_E^\ast$, and $G_C^\ast$ as~:
\begin{eqnarray}
&&\sqrt{3} \, G^+_{+\frac{3}{2} \, +\frac{1}{2}} + 
G^+_{+\frac{1}{2} \, -\frac{1}{2}} = 
I \, \frac{(M_\Delta + M_N )}{M_N \, Q_+^2} \sqrt{\frac{3}{2}} \, Q   
  \\
&&\hspace{2.5cm} \times \left\{ G_M^\ast (M_\Delta + M_N) 
+ \, G_C^\ast \, \frac{Q^2}{2 M_\Delta} \right\},  \nonumber 
\end{eqnarray}
whereas the $N \to \Delta$ form factor with two units of 
(light-front) helicity flip, corresponding with a 
quadrupole field pattern in the charge density, can be expressed as~:
\begin{eqnarray}
&&G^+_{+\frac{3}{2} \, -\frac{1}{2}} = 
I \, \frac{(M_\Delta + M_N )}{M_N \, Q_+^2} \frac{3}{4 \sqrt{2}} \, Q^2  
  \\
&& \times \left\{ G_M^\ast + 
G_E^\ast \left[ 1 - \frac{ 4 M_\Delta (M_\Delta - M_N)}{Q_-^2} 
\right] 
- \, G_C^\ast \, \frac{2 Q^2}{Q_-^2} \right\}.  \nonumber 
\end{eqnarray}

\begin{figure}[h]
\hspace{.5cm}
\includegraphics[width =7.cm]{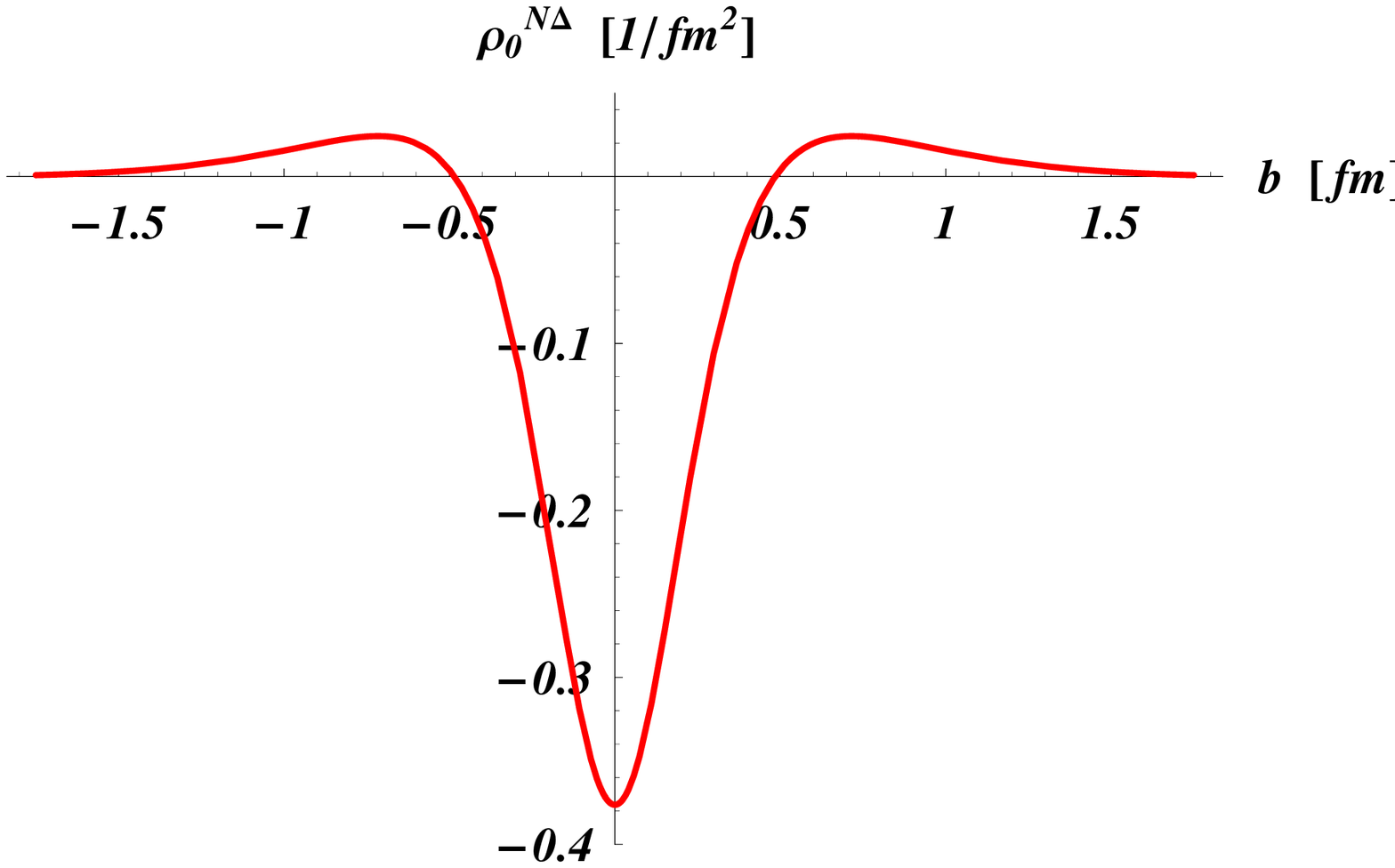}
\vspace{-.9cm}
\begin{center}
\includegraphics[width =7.cm]{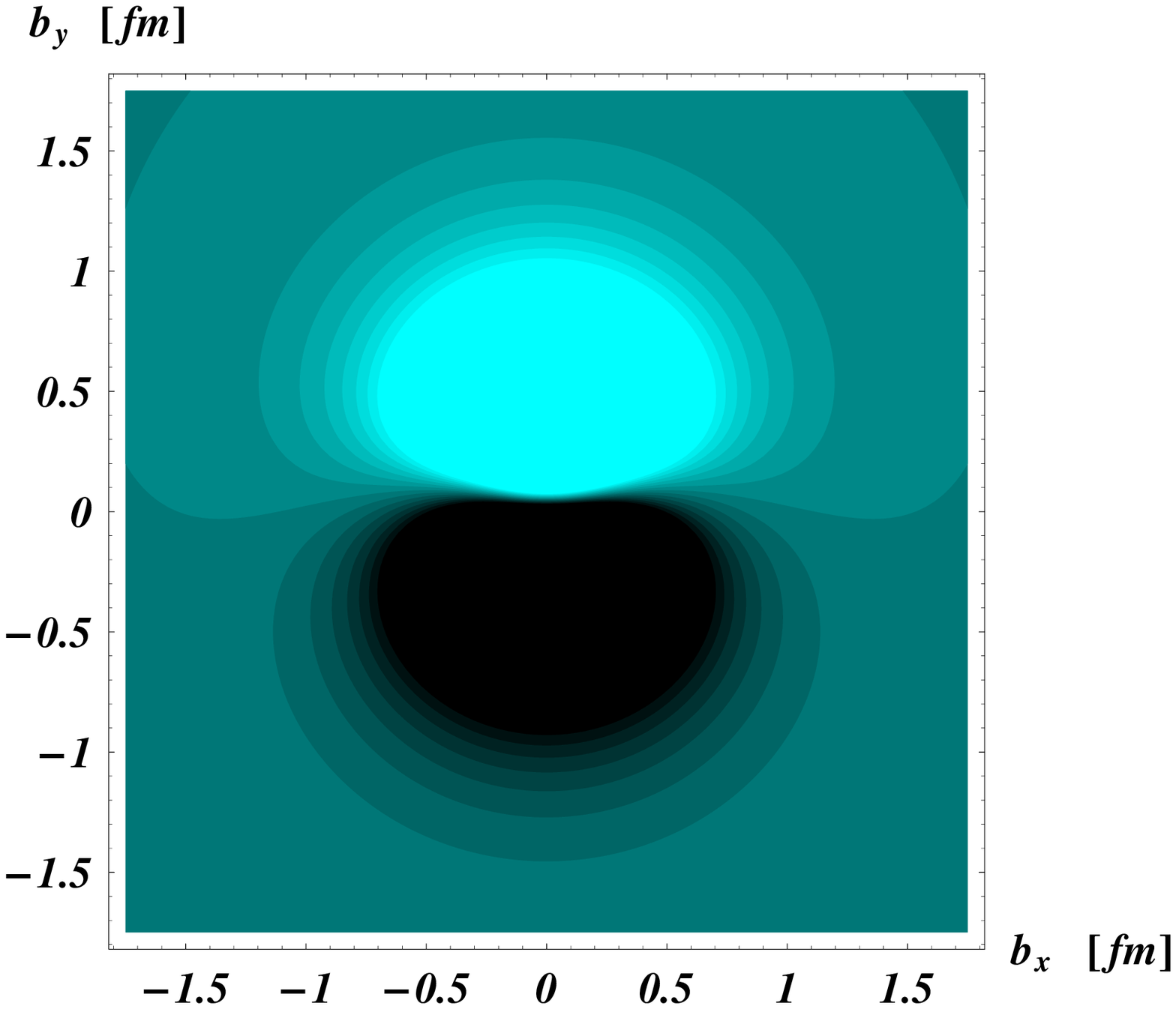}
\end{center}
\vspace{-1.5cm}
\begin{center}
\includegraphics[width =7.cm]{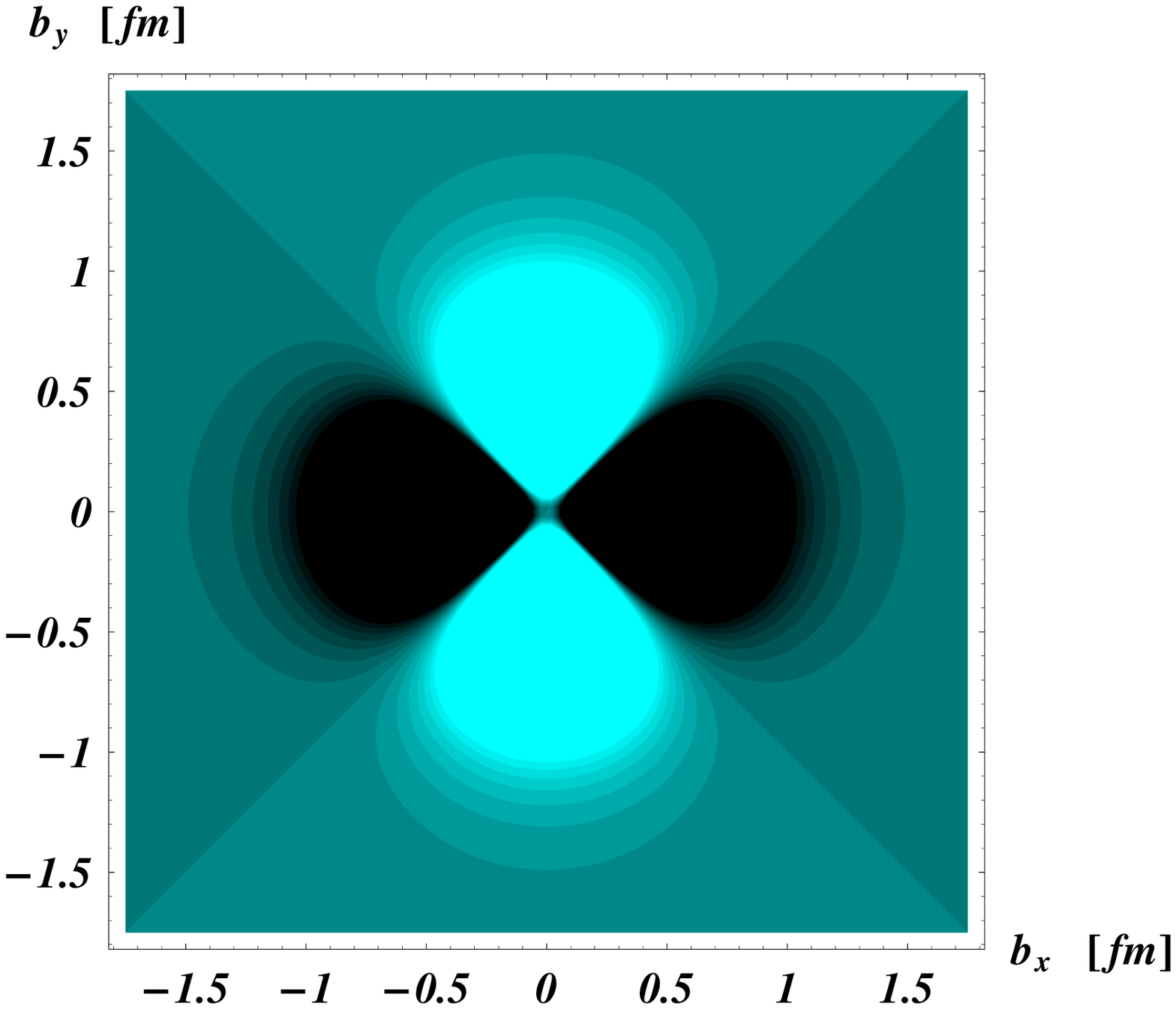}
\end{center}
\caption{Quark transverse transition charge density 
corresponding with the $p \to \Delta^+$ transition. 
Upper panel : when $N$ and $\Delta$ are unpolarized ($\rho_0^{N \Delta}$). 
Middle panel : when $N$ and $\Delta$ are polarized along the $x$-axis 
($\rho_T^{N \Delta}$). Lower panel : quadrupole contribution to 
$\rho_T^{N \Delta}$.
For the $N \to \Delta$ e.m. FFs, 
we use the empirical parameterization of 
MAID2007~\cite{MAID2007}. }
\label{fig:ndelta}
\end{figure}

We show the results for the $N \to \Delta$ transities 
densities both for the unpolarized case and for the case of 
transverse polarization in Fig.~\ref{fig:ndelta}. In this 
evaluation we use the empirical information on the 
$N \to \Delta$ transition FFs from Ref.~\cite{MAID2007}. 
One notices that the unpolarized $N \to \Delta$ transition 
density displays a behavior very similar to the neutron charge density 
(dashed curve in Fig.~\ref{fig:neutron}), having a negative interior core 
and becoming positive for $b \geq 0.5$~fm. 
The density in a transversely polarized $N$ and $\Delta$ 
shows both a dipole field pattern, 
and a quadrupole field pattern. The latter, shown
separately in Fig.~\ref{fig:ndelta}, allows to cleanly quantify 
the deformation in this transition charge distribution. 

In summary, we used the recent empirical information on 
the nucleon and $N \to \Delta$ e.m. FFs to map 
out the transverse charge densities in unpolarized and transversely polarized 
nucleons and for the $N \to \Delta$ transition.    
The nucleon charge densities are 
characterized by a dipole pattern, in addition to the monopole field 
corresponding with the unpolarized density. 
The $N \to \Delta$ transition charge density in a transversely polarized $N$ 
and $\Delta$ contains both monopole, dipole and quadrupole patterns, 
the latter corresponding with a deformation of the hadron's charge 
distribution.  

\begin{acknowledgments} 
The authors thank C.~Papanicolas and L.~Tiator for useful discussions. 
The work of C.~E.~C. is supported by the National Science Foundation
under grant PHY-0555600. 
The work of M.~V. is supported in part by DOE grant
DE-FG02-04ER41302 and contract DE-AC05-06OR23177 under
which Jefferson Science Associates operates the Jefferson Laboratory.  
\end{acknowledgments}

\end{document}